\documentclass{sig-alternate-arxiv}

\usepackage{amsmath}
\usepackage{times}
\usepackage{mathptm}
\usepackage[scaled=0.87]{helvet} 
\usepackage{graphicx}
\usepackage{balance}
\usepackage{soul}
\usepackage{color}
\usepackage{xspace}
\usepackage{url} 
\usepackage[final,inline,nomargin]{fixme}
\usepackage{clrscode} 

\newcommand{\hoot}{{\tt \#h00t}\xspace}
\newcommand{\msg}{hoot\xspace}
\newcommand{\msgs}{hoots\xspace}

\newcommand{\paragraphX}[1]{\vskip 4pt \noindent \textbf{#1} \hskip .05in}

\graphicspath{{./graphs/}}

\begin{document}

\numberofauthors{1} 
\author{
Dustin Bachrach$^{\dag}$, Christopher Nunu$^{\dag}$, Dan S. Wallach$^{\dag}$, Matthew Wright$^{\S}$\\\\
\begin{tabular}{ccc}
\affaddr{$^{\dag}$Dept. of Computer Science} & \hspace{1.2cm} & \affaddr{$^{\S}$Dept. of Computer Science and Engineering}\\
 \affaddr{Rice University} & & \affaddr{University of Texas at Arlington}\\
\affaddr{Houston, Texas, USA} & & \affaddr{Arlington, Texas, USA}  \\
\email{\{ahdustin,canunu\}@gmail.com} & & \email{mwright@cse.uta.edu}\\
\email{dwallach@cs.rice.edu} & &\\
\end{tabular}\\ \\
}


\title{{\huge \hoot}: Censorship Resistant Microblogging}

\maketitle

\begin{abstract}

Microblogging services such as Twitter are an increasingly important way
to communicate, both for individuals and for groups through the use of
hashtags that denote topics of conversation. However, groups can be
easily blocked from communicating through blocking of posts with the
	given hashtags. We propose \hoot, a system for censorship resistant
microblogging. \hoot presents an interface that is much like Twitter,
except that hashtags are replaced with very short hashes (e.g., 24 bits)
of the group identifier. Naturally, with such short hashes, hashtags
from different groups may collide and \hoot users will actually seek to
create collisions. By encrypting all posts with keys derived from the
group identifiers, \hoot client software can filter out other groups'
posts while making such filtering difficult for the adversary. In
essence, by leveraging collisions, groups can tunnel their posts in
other groups' posts. A censor could not block a given group without also
blocking the other groups with colliding hashtags.
%
%
We evaluate the feasibility of \hoot through traces collected from
Twitter, showing that a single modern computer has enough computational
throughput to encrypt every tweet sent through Twitter in real time. We
also use these traces to analyze the bandwidth and anonymity tradeoffs
that would come with different variations on how group identifiers are
encoded and hashtags are selected to purposefully collide with one
another.
%

\end{abstract}

\section{Introduction}

Recent events in Egypt, Tunisia, and many other countries have shown
that social networking sites (Facebook, Twitter, and presumably others)
played a non-trivial role in helping people organize themselves, plan
protests, and distribute videos and other news to the outside
world. Egypt was notable in that they eventually cut themselves off from
the entire Internet, in a belated and ultimately ineffectual attempt to
turn the tide. While it's difficult to draw overarching conclusions
about the centrality of social networking versus more traditional means
of communication in these important world events, it is clear that
social media played a non-trivial role. Many other countries' leaders
may well be worried of copycat revolutionaries. Other such countries may
well try to censor or otherwise tamper with their citizens' use of
social networking. To pick a current example: Syria appears to be
attempting a nationwide man-in-the-middle attack against
Facebook~\cite{syria-facebook}.

As a first step towards improving social network systems for such
environments, we seek to enable the use strong cryptographic primitives
overlaid on existing microblogging systems like Twitter, adding both
encryption and integrity to {\em tweets} (Twitter messages) among
groups. Keys must be shared to enable secure group communication, but we
should not rely on pre-arranged public key hierarchies or complex
protocols for key exchange. Users should be able to find tweets from
their group easily and then be able to receive those tweets with some
anonymity. Groups that may be targeted for their activities, even when
tweets are encrypted, should be offered {\em plausible deniability} that
they could be participating in another group instead.


To achieve these goals, we propose \hoot, a system for
censorship-resistant microblogging. Our design features an interface in
which most users never see or concern themselves with cryptographic
keys. Instead, one of our key insights is that we can overload Twitter's
{\em hashtag} mechanism as a way of deriving cryptographic key
material. \hoot can be built on top of Twitter or another microblogging
system without modifying the underlying system. Through careful design,
encrypting and decrypting {\em \msgs} (\hoot messages) and bandwidth
overhead should be acceptably small at both the server and client sides.
\hoot makes it very difficult for the censor to distinguish between the
\msgs of a group with the \msgs of a select number of other groups.


Hashtags are widely used in Twitter to label topics which others will
then subscribe to and follow. For example, most Usenix conferences adopt
the tag {\tt \#usenix}, allowing attendees to discuss the conference
with one another in real time. Political protests might end up using
several different tags (e.g., Egyptian discussions happen under {\tt
  \#tahrir}, {\tt \#jan25}, {\tt \#25jan}, and {\tt \#egypt}, among
others). Hashtags searches are generally case-insensitive.

Some tags have staggering volumes of messages. To pick a notable
example, pop singer Justin Bieber asked his roughly 9 million followers
to discuss his movie, {\em Never Say Never} using the hashtag {\tt
  \#nsn3d}. At its peak, roughly 1\% of Twitter's traffic mentioned this
tag\footnote{Statistics via Trendistic, Topsy, and HashTracking.}. Since
the movie's release in February 2011, there have been roughly 164
thousand tweets using {\tt \#nsn3d}, an average of 1.6 per minute with
significantly higher peaks. 500 recent tweets on this hashtag generated
346 thousand impressions, reaching an audience of 212 thousand followers
within a 24 hour period (measured in mid-April 2011).

Of course, not all tags are as popular. We will show later, in
Section~\ref{sec:experiments}, that hashtag usage follows a power-law
distribution; a small number of hashtags are incredibly widely used and
large numbers of tags are used very rarely or only once. We would like
to design a system that can leverage these communications to create
cover traffic for other, more sensitive messages, but without simply
reusing the popular hashtag for other content. This will require
converting hashtags into cryptographic keys and arranging for them to
collide in some fashion, such that a query for {\tt \#nsn3d} and a query
for a more sensitive tag are indistinguishable to an observer, thus
providing some measure of deniability to group subscribers (``Protests?
I'm just a fan of Justin Bieber!''). We also need to give some amount of
control to the organizers of the sensitive communications, allowing them
to select any popular hashtag with which they might prefer to collide.

Ultimately, we see two main paths to designing our system. One option would be to send encrypted messages that include the real {\tt \#nsn3d} tag, perhaps engineering some sort of steganographic process that tries to hide the plaintext within messages that are statistically similar to other posts from Bieber's fan, but it seems inappropriate to produce false messages like this. The other possibility is to imagine that {\em all} Twitter messages are encrypted in a uniform way, where knowing the plaintext of the hashtags would enable the decryption of a message. (It's easy to see a proxy server, of some sort, providing an ``encrypted'' interface to Twitter in this fashion
.) This is the design we chose to pursue. In this setup, we can encrypt and MAC every message with a random session key, which can be decrypted if the user knows the proper hashtag. ``Encrypted'' hashtags can also be generated by hashing the plaintext hashtags and truncating those hashes.  (We address this unwieldy vocabulary when we present our design in Section~\ref{sec:design}.) Consequently, two different plaintext hashtags can collide with each other with a probability related to the number of bits in the truncated hash.

\section{Threat model and system goals}

In this section, we briefly outline our threat model and then describe
the goals of our design.

\subsection{Threat model} \label{sec:threat}
\hoot is designed to provide censorship resistance against an adversary
who can observe all \hoot traffic and can block any tweets it
chooses. In practice, the adversary may only block tweets being received
by a subset of users, but this does not affect our model. The adversary
seeks to identify and block tweets discussing a small set of topics,
based on the identity of the sender, keywords in the content, or the
hashtag itself. The adversary blocks SSL access to the service or acts
as a man-in-the-middle to enable eavesdropping and selective blocking.
However, our adversary does not want to disable the entire service, nor
does our adversary want to disable popular but innocuous discussion
threads; such crude censorship might then generate additional unrest in
the population.

We argue that some governments will be interested in this level of
censorship, in which the microblogging service is allowed in a
restricted fashion. We note that some countries are currently censoring
services such as Facebook and Twitter in their entirety. As
microblogging services become an increasingly important form of
communication, however, we believe that most countries will find
complete censorship of these services to be incompatible with operating
in the modern world. It is possible that blocking all microblogging
services could be seen as heavy-handed as blocking email would be today.
Furthermore, Kuppusamy and Shanmugam~\cite{ict-economy} showed that
information technology and communication leads to greater economic
growth. Broad censorship reduces the utility of the entire system, which
is presumably used for economic activities more valuable than discussing
teenage pop stars.


The adversary has moderate computing power and can perform a brute force
key space search over a reasonable space. We describe
how to address stronger adversaries in Section~\ref{sec:security}.

We assume that the \hoot server does not cooperate with the adversary. We
also assume that the adversary has no insiders in the group who leak the
group's secrets. Likewise, we assume the adversary has no avenues
to attack users, such as setting up a covert keylogger
on a group member's machine or coercing a group
member to divulge a secret. We concede that such attacks would effectively undermine \hoot's censorship
resistance. Note that this serves as a practical bound on any attack; if
it would be easier to establish an insider in the group or subvert one
of the client systems than to perform a computational attack, we will
consider our design to be successful.
We further discuss mitigations to such attacks in Section~\ref{sec:discuss}.

\subsection{System goals} \label{sec:goals}
Our goal is to allow a user to send a secure message to a private group
of individuals, allowing only the group members to read the plain-text
message, and to accomplish this with a user interface that looks and
feels much like the vanilla Twitter interface. Ultimately, this creates
a variety of constraints and challenges.

\begin{description}

\item[Simple key distribution.] To make the system as easy to use as
  possible, keys should be simple to create, distribute, and use. We
  therefore rule out any cryptographic key hierarchy such as a public
  key infrastructure or PGP/GPG key signing parties. Instead we wish
  to have keys that can literally be passed via
  word-of-mouth, from person to person in the group. We propose
  to derive keys from the group's plaintext hashtag, which effectively
  serves as the membership password. When a user subscribes to a given plaintext
  hashtag, she inputs the hashtag into her \hoot client, and the client
  derives the necessary keys.

\item[Confidentiality of tweets.] Since the plaintext hashtag is being
  used to generate encryption keys, it should have sufficient entropy to
  protect against dictionary attacks and brute force. This is in tension
  with our desire to have the plaintext hashtags be easily memorized and
  shared between users, ideally by voice alone.

\item[Censorship resistance and denial of service.] While we do not
  attempt to defeat censorship of the \hoot service in its entirety, we
  seek to defeat attempts to censor specific groups and keywords. Perng
  et al.~\cite{perng05revisited} define {\em censorship susceptability}
  as the probability that the adversary can block a targeted message
  while allowing at least one other message to be received. This is a
  difficult requirement to meet in our system. We instead aim to allow
  only {\em heavy-handed censorship}, which we define as censorship of a
  group only through censorship of multiple, unrelated groups. By
  censoring these groups together, the adversary lowers the utility of
  the system as a whole. This is similar to the resistance provided by
  document-based systems like Tangler and
  Dagster~\cite{tangler,dagster}.

\item[Recipient anonymity.] To achieve censorship resistance, we rely on
  being able to protect recipient anonymity. Adapting the definition
  from Pfitzmann and Hansen~\cite{terminology}, we require that the
  recipients of the tweets, i.e. the group members, not be identifiable
  from among a larger set of possible recipients. \hoot makes this
  possible by mapping plaintext hashtags, which identify the groups, to
  short hashtags that can be made to collide with those of other
  groups. All of the recipients in all groups with colliding hashtags
  form a recipient anonymity set, with the tweets from other colliding
  groups providing cover traffic. \hoot can also be said to provide
  {\em subscriber anonymity}, as introduced by Mislove et al. in their
  description of AP3~\cite{ap3}. Hordes~\cite{hordes} and P5~\cite{P5}
  have similar requirements. The main additional feature of subscriber
  anonymity over recipient anonymity is that the act of subscribing
  should not reveal information that could be used to break recipient
  anonymity.


\item[Recipient deniability.] If a \hoot user is under physical threat to
  reveal what hashtags she subscribes to, it's important that she can
  offer a convincing lie. Through careful selection of groups with
  colliding hashtags, she could name the hashtag of an innocuous group
  that could reasonably be of interest to members of the target group.
  Suitable choices can include trending topics (e.g., the Justin Bieber
  movie hashtag {\tt \#nsn3d}), socially-appropriate discussion groups
  (e.g. {\tt \#Bible} or {\tt \#Quran}), or topics that related to
  other innocuous professional or personal interests.

\item[Sender anonymity or deniability.] Along those lines, we can only
  provide limited protection to a sender. A sender should gain some plausible
  deniability against a passive attacker, in that she could be
  tweeting about any possible topic that collides with her post.
 If, however, we must resist physical attacks against a sender,
coercing them to decrypt a posted message, our core \hoot design
will not protect them.
Instead, message senders who need to remain anonymous or who require the
ability to deny having posted a given message must use external means,
such as Tor, to connect to the \hoot service for posting messages. (If a
decentralized or P2P transport mechanism was used for microblogging,
like BirdFeeder~\cite{sandler09}, such a system could be extended to have
anonymous posting features. For this specific research, we are
generally targeting a centralized service more like Twitter.)

\item[Replay attacks.] It's possible that a malicious user, or even a
  malicious microblogging service, could not only remove messages but
  could also replay old messages, possibly with telling side effects
  (e.g. ``Meet in the town square at noon.''). We must have 
  mechanisms to reject duplicates.

\item[Statistical and traffic analysis.] Even if the adversary cannot
  decrypt messages, it may be able to learn things by scanning large
  populations of \msgs. While we make no explicit attempt to hide who
  the sender of a message might be (see ``sender anonymity,'' above), we
  do want to provide a strong degree of resistance to traffic analysis
  that might otherwise bind senders to receivers. Our system should make
  it difficult or impossible for observers to reconstruct the social
  graph.

\item[Secret informers and coerced users.] If a group member, whether
  sender or recipient, is an insider for the adversary or if the
  plaintext hashtag is stolen through a keylogger or coersion, the key
  is compromised and the group's messages will become readable and
  censorable to the adversary. While we cannot stop such attacks, we
  clearly need some form of key agility, to allow group organizers to
  distribute new hashtags to replace older, compromised hashtags.

\item[Compatibility.] We want to ensure that \hoot can be layered on top
  of Twitter, using existing Twitter mechanisms to search for and follow
  desired messages. We also must ensure that real Twitter users could
  incrementally migrate to using \hoot  as a service above the existing
  Twitter. To that end, we must demonstrate that we can implement
  efficient proxy servers, converting Twitter to \hoot to bootstrap an
  effective \hoot rollout.

\end{description}

One goal that we do not seek to achieve is {\em membership concealment},
which Vasserman et al. define as hiding the fact that the members are
participating in the system~\cite{vasserman09mcon}. The rationale for
membership concealment is that employing a tool designed to circumvent
the censor will draw unwanted attention to the user. By building \hoot
on top of a popular communications medium (Twitter), ideally with many
groups using \hoot in place of normal hashtags, we argue that \hoot
could be deployed in such a way that users are typically not aiming to
circumvent censorship. Since \hoot also provides message privacy,
authentication, integrity, and receiver anonymity, groups have other
reasons to use it instead of plaintext tweets besides censorship
resistance. If \hoot is widely adopted over Twitter for typical group
communication, a censor that blocks other systems for
censorship-resistance (such as Tor bridges), might not be willing to
block all \msgs.

Whether this is the case in the real world is hard to discern. China
appears to have blocked iTunes for about 10 days in 2008 due to a
pro-Tibet album; however, it restored service while blocking the album
page itself~\cite{china-itunes}. Further, there was not a China-specific
iTunes service at the time. If the colliding groups are popular in the
country under censorship, then blocking results in the censorship being
widely seen inside the country and raises awareness of
censorship. Google discloses to users in China when their searches have
been modified\footnote{See
  \url{http://googleblog.blogspot.com/2006/02/testimony-internet-in-china.html}},
providing a similar type of awareness.

\section{Design \& Security Analysis}
\label{sec:design-sec}

In this section, we describe the \hoot protocol and analyze its security.

\subsection{Design}
\label{sec:design}

We now describe \hoot in detail. After giving a brief overview of the
\hoot  protocol, we describe how hashtags are generated to provide
collisions with other groups and how the message header and body are
constructed to enable efficient searching.

\paragraphX{Protocol overview.} A complete \msg consists of a header and
a message body. The header contains a group identifier (a Twitter-style
hashtag), an encryption key and a MAC key, both encrypted with a session
key, and finally a MAC over the ciphertext of the message (see
Figure~\ref{fig:hoot-structure}). As in Twitter, \msgs do not name their
recipients. Anyone who knows the secret hashtag associated with a \msg
can decrypt and read the message as well as validate its integrity. We
also need an efficient discovery mechanism.  Rather than attempting to
treat every message posted to Twitter as a potential group message, and
thus being required to fetch and attempt decryption of every single
message, the \hoot protocol places an identifier into every \msg as a
hashtag so a fellow group member can simply search for the identifier to
see all potential messages. With a constant group identifier, readers
can also publicly follow that identifier like any other hashtag on
Twitter.

\paragraphX{Group identifiers.} To create a hashtag for use as the group
identifier, \hoot derives a fixed-length bitstring from the secret
hashtag. We must do this in such a way as to give an attacker no
information about the shared secret itself. A cryptographic hash
function serves this purpose well, but makes brute force very easy. We
recommend a more expensive key derivation function, such as scrypt,
which works well against brute force even against optimized
hardware~\cite{scrypt}. Percival estimates that it would require
\$610,000 of specialized hardware to crack an 8-character,
scrypt-secured password that included only lower-case letters.
We call the secret hashtag a \textit{plain tag}, which is comparable to
a normal Twitter hashtag, though it should have enough entropy to
prevent the adversary from guessing it. The result of the key derivation
function \textit{H} is referred to as the \textit{long tag}, i.e.:
$\id{LongTag} \leftarrow H\left(\id{PlainTag}\right)$.

The \hoot protocol could simply use the long tag as an identifier, but
this choice leads to several problems. First, to achieve our design goal
of keeping identifiers short and to fit within Twitter's 140 character
limit, it is less than ideal to use the full output of a key derivation
function (e.g. 128 bits). Secondly, a good key derivation function, much
like a cryptographic hash function, produces virtually no collisions for
reasonable numbers of groups. As described in Section~\ref{sec:goals},
we propose that different groups' identifiers collide with each other
for recipient anonymity and plausible deniability.

To generate a collision, we need to shorten the long tag, generating a
\textit{short tag} of $k$ bits. The short tag will, by design, induce
collisions between unrelated plain tags. The shorter the short tag, the
higher the collision rate will be and the less sure an observer can be
as to what topic a \hoot reader is actually following. With this greater
anonymity comes more computational work: since more group messages will
now belong to the same identifier, a follower must download and decrypt
more messages to find the desired ones.

Given a consistent system-wide short tag length, a group can choose a
tag that will collide with a popular tag, allowing for a predictably
high amount of cover traffic as well as providing a cover story for
followers of that tag.
%
\begin{figure}
\begin{codebox}
\Procname{$\proc{Find-Tag}(\id{prefix}, \id{target}, N, k):$}
\zi \For $i \gets [0,N)$, in random order
\zi \Do
\zi $\id{PlainTag} \gets  \id{prefix}.\id{suffix}$
\zi $\id{ShortTag} \gets H(\id{PlainTag}).\func{bits}(0 \ldots k-1)$
\zi \If $\id{ShortTag} = H(\id{target}).\func{bits}(0 \ldots k-1)$
\zi \Then $\func{return} (\id{PlainTag} , \id{ShortTag})$
\zi \End
\End
\end{codebox}
\caption{Pseudocode for tag collision searching.\label{fig:find-tag}}
\end{figure}

%
This algorithm searches for a tag collision, where the \id{PlainTag}
suffix is a number between 0 and $N$, and the \id{ShortTag} is $k$ bits
long.  What should be reasonable values for $N$ and $k$?

$k$ determines the length of the \id{ShortTag}. As discussed above, the
value for $k$ trades off anonymity versus search overhead for a
receiver. $k$ will likely need to be a constant shared widely across the
space of \hoot users.

$N$ is bounded by how large a \id{PlainTag} string can be reasonably
passed among potential \hoot participants. If the communication of the
\id{PlainTag} must happen by word of mouth, $N$ will be bounded,
perhaps, by the number of digits that can be memorized by most humans
(so if humans can remember around seven decimal digits~\cite{miller56},
then $N$ would be $10^7$). Equivalently, we could search over some other
memorizable namespace with suitably high entropy, like a short string of
characters found on a keyboard.  Regardless, the group creator would use
a \proc{Find-Tag} procedure (see Figure~\ref{fig:find-tag}) to search
over all possible suffixes to identify collisions. Note that the search
should be done randomly, rather than in-order, to increase the
attacker's difficulty in conducting brute force attacks. Also not that
process is only necessary once, when a tag is first created.

To further increase the entropy of the plain tag, we can imagine a
number of options that would still be amenable to human
memorization. For example, the short tag's prefix could be chosen
randomly from a large dictionary or replaced with a full phrase. NIST
estimates a 40-character pass phrase with no checks or restrictions to
have about 56 bits of entropy~\cite{nist}. If we were willing to relax
our desire to have human-memorizable plain tags, then the whole plain
tag could be selected at random. Certainly, this yields excellent
resistance to brute force searching attacks, but it also creates
additional complexity for organizers wishing to prevent leaks, since
these plain tags will need to be written down or saved and shown on a
mobile device.

\paragraphX{Message header and body.}  
In addition to the \id{ShortTag}, the header contains a pair of
session keys
for message body encryption ($k_{\func{enc}}$) and integrity
verification ($k_{\func{mac}}$).

For every \msg, these session keys are randomly generated. Since we
intend to use efficient symmetric key ciphers and hash-based message
authentication functions.
The session keys
are then encrypted with a {\em tag key} derived from the long tag, using
different bits than the $k$ bits used when deriving the
short tag. Given a long tag of 160 bits, if we assume half of those bits
are used in the short tag, the remaining 80 bits give
us $2^{80}$ possible keys that an attacker must potentially brute
force, which is certainly greater than the entropy in the
plaintext tag. (In Section~\ref{sec:experiments}, we flesh this out
in more detail.) Of course, if we ever reached a point where the
encryption and MAC session keys required more
bits than we can get from carving up the long tag, we could always
use the long tag to initialize a suitably strong pseudo-random number
generator, getting us all the derived bits we might ever want.

%

\begin{figure}
\begin{eqnarray*}
M & \leftarrow & \mathrm{plaintext\ message,\ including} \id{PlainTag}
\\
\id{LongTag} & \leftarrow & H(\id{PlainTag}) \\
\id{ShortTag} & \leftarrow & \id{LongTag}.\func{bits}(0 \ldots k-1) \\
k_{\func{tag}} & \leftarrow & \id{LongTag}.\func{bits}(k \ldots) \\
k_{\func{enc}}, k_{\func{mac}} & \leftarrow & \id{random bits} \\
C & \leftarrow & E_{k_{\func{enc}}}(M) \\
\id{HooT}  & \leftarrow &  \left(\id{ShortTag}, E_{k_{\func{tag}}}\left(k_{\func{enc}}, k_{\func{mac}}\right), \func{MAC}_{k_{\func{mac}}}(C), C\right)
\end{eqnarray*}
\caption{Structure of a \msg.\label{fig:hoot-structure}}
\end{figure}

So far, we have specified a \msg structure with exactly one plain tag.
This technique can easily be generalized to support multiple plain tags. For
each one, a separate long tag can be generated, resulting in multiple
tag keys ($k_{\func{tag}}$), each of which is used to encrypt the same
session keys. The final \hoot would have multiple short tags and multiple
encryptions of the session keys, but only one ciphertext message
payload.

\begin{table*}
\caption{This table shows how several tweets might be converted to
  \msgs, showing the long tag, the short tag, and the final \msg. The
  fourth message in this list demonstrates how a group could take
  advantage of the \hoot system to collide its \msgs with those of an
  unrelated tag used for non-controversial messages.
\label{tab:process}
}
\begin{center}
    \begin{tabular}{ l  l  l  l  l }
	 & Tweet & Long Tag & Short Tag & Hoot \\ \hline
	1 & Its all bout the {\bf \#bieber} 100\%Belieber                                 & {\tt 9txrq71tfn8} &  {\tt 9tx} & {\tt \#9tx Xrtfn}... \\
	2 & Don't be a drag; just be a queen whether you're broke or {\bf \#CharlieSheen} & {\tt 7prQnd121f2} & {\tt 7pr} & {\tt \#7pr n771r}... \\
	3 & {\bf \#free-egypt} We'll meet at the usual, 11pm.                             & {\tt 2p7rtfx9pa1} & {\tt 2p7} & {\tt \#2p7 pp76a}... \\
	4 & {\bf \#free-egypt-9rqt} We'll meet at the usual, 11pm.                        & {\tt 9tx79srpLtt} &  {\tt 9tx}  & {\tt \#9tx 18yyQ}... \\
    \end{tabular}
\end{center}
\end{table*}

For illustration, Table~\ref{tab:process} shows how a few plain-text
tweets might be converted into their corresponding \msgs. The first two
messages are regular tweets from popular hashtags: \#bieber and
\#CharlieSheen. The third is a \msg where receiver anonymity is
critical, but it's short tag, {\tt \#2p7}, does not collide with
anything else, and thus subscribers to {\tt \#2p7} might risk
discovery. The fourth message shows how the same group might alter their
plain tag so they can deliberate collide with {\tt \#bieber}, which maps
to the same short tag ({\tt \#9tx}).

\subsection{Security analysis}
\label{sec:security}

Based on the threat model defined in Section~\ref{sec:threat} and the
system design goals described in Section~\ref{sec:goals}, we now analyze
the security of the proposed \hoot protocol.

\paragraphX{Message security.}  
We begin with a brief analysis of the security of the message protocol
itself.

First, note that the session keys are generated randomly and
independently for each \msg. Consequently, two identical plaintext
messages will have different ciphetexts. If the encryption scheme in use
requires an initialization vector (e.g., CBC mode), this could be safely
included in the message header. For other encryption schemes, such as
counter mode, no IV is necessary and the randomness of the key will
ensure the non-determinism of the ciphertext.

Message integrity is validated with a symmetric-key message
authentication code such as HMAC-SHA1. Because the MAC is computed over
the ciphertext, and the MAC key is generated at randomly and
independently from the encryption key, the MAC leaks absolutely no
information about the plaintext. The MAC verification process also
serves the purpose of identifying whether a prospective \msg matches the
plain tag in question (for which multiple other plain tags will collide
in the short tags), or whether a message is irrelevant to the user's
plain tag search query and should be dropped.

Replay attacks can be defeated by treating the session keys
($k_{\func{enc}}$, $k_{\func{mac}}$) as nonces. It's highly unlikely
that two different \msgs will share the same session keys.


\paragraphX{Brute force attacks.}  
Provided an attacker knows a targeted group's prefix and the alphabet
out of which they generate the suffix, our scheme is amenable to brute
force searching attacks. Table~\ref{tab:hps} shows how fast a modern
computer can decrypt \msgs: between $2^{17}$ and $2^{18}$ per
second. Clearly, plain tags must be selected with far more entropy than
this. If the attacker has $2^{10}$ CPU cores and we want a plain tag to
survive one week of analysis (just over $2^{19}$ seconds) before a plain
tag is ``burned'' and needs to be replaced, then plain tags would
require a minimum of 47 bits of entropy (e.g., 15 decimal digits).

If we limited plain tags to a word from a reasonably large dictionary
(40,000 entries) plus 7 decimal digits, then we only get 38.5 bits of
entropy.  Our hypothetical attacker with $2^{10}$ CPU cores could brute
force a plain tag in 20 minutes.

If a plain tag was composed of two dictionary words and 7 decimal
digits, yielding 53.8 bits of entropy, then our hypothetical attacker
would need over two years of computation to brute force a plain
tag. While this is certainly pushing the boundaries of what might be
memorable without being written down, it's not inconceivable.

\paragraphX{Traffic analysis and adaptive censorship.}  
Consider the case where an attacker can see what queries are
subscribed to by each individual user within their country. The
attacker suspects that there is hidden traffic on a particular short
tag, based on the prevalence of queries for it, so the attacker
proceeds to twist some arms and finds what appears to be a
sudden and inexplicable rise in domestic fandom for a teenage pop star
from a foreign country.

Is this falsifiable? Ironically, the locals who have chosen the foreign
pop star for their cover traffic can best cover themselves by immersing
themselves in the pop star's oeuvre. Still, the pop star's genuine
traffic is no secret. The attacker could censor the short tag, in its
entirety, accepting the false positives and causing outrage among the
pop star's true fandom. Alternately, the attacker could censor the \msgs
on the short tag that {\em do not match} the pop star's known plain
tag. Of course, this would also have false positives with legitimate and
innocuous traffic, but it would definitely force the organizers to shift
their traffic to a different short tag, creating something of a game of
cat-and-mouse.

We note that the pop star could choose to surreptitiously help his
overseas ``fans'' by regularly adding new plain tags under which he
implores his genuine fans to discuss new topics (e.g., his new
haircut, his new hit single, his guest appearance on a talk show, and
so forth). The local organizers could then take advantage of this by
running \proc{Find-Tag} to discover tags that collide with each
one. No actual communication between the foreign pop star and his
local ``fans'' would ever be necessary.

\section{Implementation}
\label{sec:implementation}

In this section we describe our prototype \hoot implementation, which we
use for performance experiments (see
Section~\ref{sec:experiments}). Additionally, our discussion in this
section helps to illustrate the design choices and trade-offs available
in the \hoot approach.

\paragraphX{Generating a Hoot}
The \hoot protocol allows for a variety of different encryption,
hashing, and message authentication schemes. Our prototype client,
implemented in Python using the
PyCrypto\footnote{\url{http://www.dlitz.net/software/pycrypto/}}
library, takes the following steps to construct a \msg:
\begin{itemize}
\item We generate a long tag by taking a SHA-1 hash of the plain
  tag.\footnote{Unfortunately, we did not have time to implement and
    test with scrypt. Scrypt runs much slower, but can be tuned for
    different trade-offs of speed and security~\cite{scrypt}.}  This
  provides 32 bits for the short tag and 128 bits for the tag key,
  $k_{\mathrm{tag}}$, to encrypt the session keys.
\item Using PyCrypto's cryptographically-strong random number
  generator, we generate 
  a random 128-bit encryption key, $k_{\mathrm{enc}}$, and a random 128-bit MAC key,
  $k_{\mathrm{mac}}$. These keys are
  concatenated together and then encrypted with the tag key using AES in
  counter mode with a fixed initial counter of $0$. (The randomly
  chosen encryption key for AES ensures a suitable level of
  non-determinism in its ciphertext.)
\item We encrypt the plaintext message with $k_{\mathrm{enc}}$, again using AES in
  counter mode, and use $k_{\mathrm{mac}}$ to generate an HMAC-SHA1
  message authentication code over the encrypted plaintext of the
  message.
\item We print the \msg, consisting of a {\tt \#} symbol, the short tag,
  a space, the encrypted keys, the HMAC digest, and the ciphertext.
\end{itemize}

(We describe a prototype of \proc{Find-Tag} in
Section~\ref{sec:collider}.)


\paragraphX{Message length.}  
We wish to render \msgs in a format that can be transmitted via
Twitter. The primarily difficulty we face is Twitter's 140 character
limit. We must also ensure that the short tags are rendered in standard
Twitter hashtag format (i.e., preceded with a {\tt \#} character and
followed by whitespace) such that standard Twitter searching mechanisms
will efficiently find them.

Interestingly, Twitter has a very broad definition of a character. Based
on our testing, we believe that Twitter limits tweets to 140 Unicode
(UTF-8) glyphs. While we could certainly take advantage of this to
squeeze the longest possible \msgs into a single tweet, particularly if
we were willing to restrict plaintext messages to 7-bit ASCII, we chose
not to pursue this for our initial implementation. Instead, we went with
a standard Base64 encoding (the letters A-Z, a-z, 0-9, +, and /),
yielding only six bits per glyph.

Assuming a single short tag of two Base64 glyphs (12 bits), the maximum
plain text message with our prototype implementation would be 31
single-byte characters. (We would need 79 glyphs to represent the
message header, including one short tag, leaving 61 glyphs for the
message, which could then be at most 45 bytes of plain text.)

If, however, we were to implement a more efficient Unicode packing, we
could certainly do much better. UTF-8 allows for just over 1 million
values in a single glyph, not all of which are currently in use\footnote{Wikipedia has a
  reasonably good discussion on this topic:
  \url{http://en.wikipedia.org/wiki/UTF-8}.}. As such, a Unicode packer
should be able to achieve 20 bits per glyph. With this, the entire \hoot
header, with one short tag, could be encoded in 26 glyphs, leaving 114
glyphs for the ciphertext. For users used to Twitter's 140-character
length restriction,
this is likely to be reasonable (moreso if they limit themselves to
7-bit ASCII characters). Of course, we could also build any
compression scheme into the \hoot protocol, perhaps adding a 
few bits to the header to indicate a language group, and thus
initialize the compression system with a corpus of common words
and phrases. This would radically improve the efficiency of the
compression scheme and thus the amount of data that could be encoded
in a \hoot, while still respecting Twitter's 140-character maximum.

We note that Twitter is not the only microblogging system that we could
leverage for \hoot traffic. \hoot could just as easily be built atop
Google Buzz, which doesn't have Twitter's 140 character
limit.\footnote{The 140 character limit is an artifact of Twitter's
  original intent to support SMS cellular telephone messaging as a way
  of delivering tweets. This unfortunate design decision was originally
  made in 1985 and still haunts us today~\cite{latimes-char160}.}

\section{Experiments}
\label{sec:experiments}

In this section, we describe the results of our analysis of the \hoot
system with the 2008 Twitter dataset collected by Sandler and
Wallach~\cite{sandler09}. This data contains 10,766,525 tweets with
255,833 hashtag references. Additionally, we show the results of
experiments using our prototype to examine the computational overhead of
implementing \hoot over all Twitter traffic.

\subsection{Cover traffic}
We first show that Twitter groups provide good possibilities for cover
traffic that \hoot can leverage to hide groups seeking plausible
deniability.

\begin{figure*}[t]
\begin{minipage}[b]{0.48\linewidth}
\centering
\includegraphics[scale=.5, viewport=0cm 0cm 16.6cm 13.6cm]{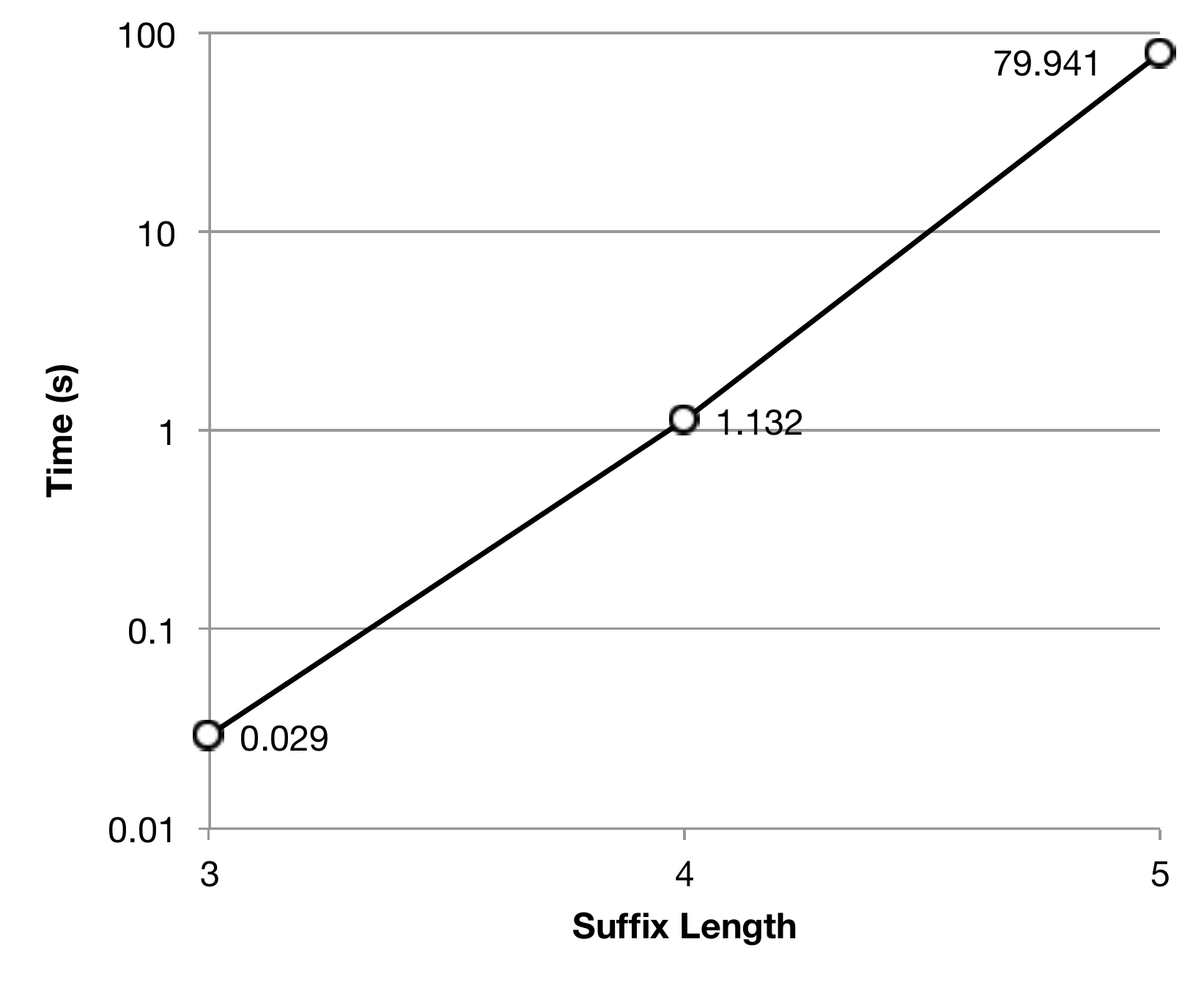}
    \caption{Runtime for the collider to search for all matching tags with
  suffixes of length $L=3,4,5$ base-64 digits on a PC with dual quad-core Intel i5
  processors.}\label{fig:collider-times}
\end{minipage}
\hspace{0.5cm}
\begin{minipage}[b]{0.48\linewidth}
\centering
\includegraphics[scale=.5, viewport= 0cm 0cm 16.6cm 12.9cm]{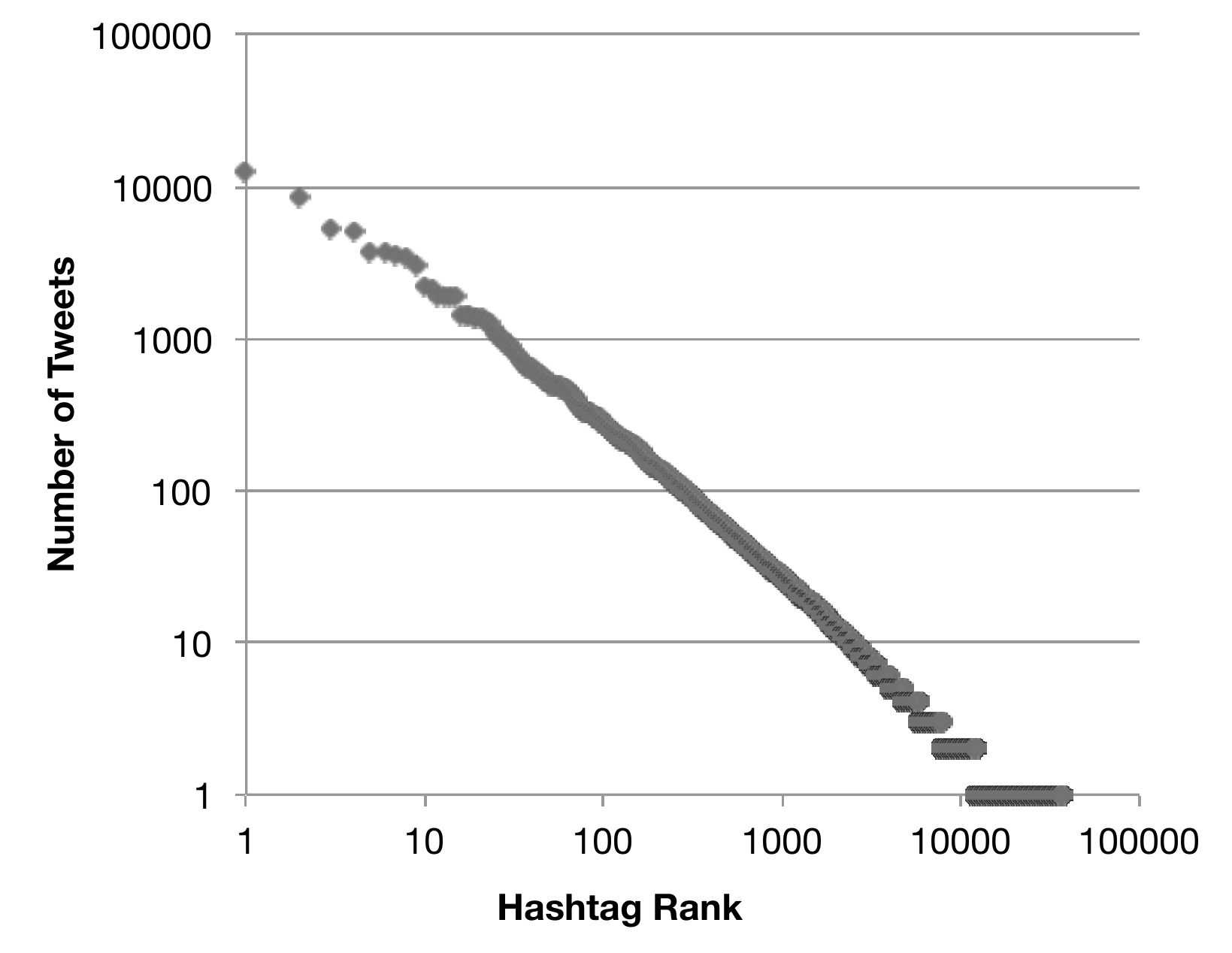}
\caption{2009 Twitter hashtag activity distribution on a log-log
  scale.\vspace{0.4cm}\label{fig:hash-dist}
}
\end{minipage}
\end{figure*}

In Figure~\ref{fig:hash-dist}, we show the distribution of 
tweet volume for each hashtag in the 2008 dataset, ordered by activity
volume in a log-log scatter plot. The
distribution appears to follow a standard power law distribution, with a
few very active hashtags and many hashtags with few tweets. A large
cluster of hashtags appear only once in our dataset. 

This distribution shows us that there is a large spectrum of subscriber
anonymity set sizes that can be leveraged by \hoot. To get a high
degree of anonymity, a group organizer can choose a plain tag whose short tag
collides with a popular tag (and, thanks to the power-law
distribution, we can be confident there will always be a reasonable
distribution of plain tags, with varying popularity, to choose from).
Among other benefits, a group organizer has the ability to dial in
pretty much any amount of cover traffic for their group.



\subsection{Finding tags}
\label{sec:collider}

To explore the feasibility of finding a plain tag that collides with an
existing short tag, we built a tool called the {\em collider} in C and
OpenMP~3.0, using the open source CommonCrypto\footnote{\url{http://www.opensource.apple.com/source/CommonCrypto/}} library provided
by Apple. The collider implements the \proc{Find-Tag} algorithm
described in Figure~\ref{fig:find-tag} and is trivially parallelizable
by partitioning the search space.

As expected, the runtime scales exponentially with the length of the
desired collision.  Performance is limited by the performance of SHA-1
of our computer--- roughly 1.9 million ($~2^{21}$) hashes per second per
core.
Consequently, finding a collision in a short tag that's only three
Base64 digits (18 bits) takes a fraction of a second. Finding a
collision in four digits takes only a few seconds. Even with our
parallelized implementation, it is currently infeasible for a single PC
to do an exhaustive search for a suffix greater than six Base64 digits
(36 bits) in less than a day. Note that performance would be slower if
scrypt or any other slow key derivation function was used in place of
SHA-1.

\subsection{Overall \hoot performance}

Adoption of \hoot requires that the overhead for encryption and
decryption be minimal. For example, one path to adoption would have
Twitter or another microblogging service offer \hoot semantics over
their entire existing service. We thus consider computation overhead in
the context of encrypting {\em all} Twitter traffic as a worst case. In
March 2011, Twitter stated that the site receives 140 million tweets per
day or 1620 tweets per second on
average\footnote{\url{http://blog.twitter.com/2011/03/numbers.html}}. Twitter
also stated that the maximum number tweets per second ever was
6939. These numbers act as rough upper bounds to the number of hoots per
second the system would need to keep up with.

\begin{table}
\caption{\hoot computation rate for encryption and
  decryption.\label{tab:hps}}
\begin{center}
    \begin{tabular}{ l  r }
	Action & Average hoots per second \\ \hline
	Encryption & 3610 \\
	Decryption & 15590 
    \end{tabular}
\end{center}
\end{table}

To study the amount of computation required to support \hoot over
Twitter, we modified our Python script to perform the encryption process
500,000 times running on a MacBook Air with a 1.86~GHz Intel Core~2 Duo
using Base64 encoding to demonstrate how the performance would be on a
typical end user's laptop. This also gives us a lower bound for
performance on modern hardware. We also independently performed the
decryption process 500,000 times.  Table~\ref{tab:hps} demonstrates that
the computational overhead for \hoot cryptography is negligible for
Twitter; the average computational load can be handled by a single
computer and the peak load might require only two computers. Clearly,
the issue for Twitter or a comparable service wouldn't be cryptographic
costs, it would be bandwidth. That overhead would be entirely dependent
on the selection of the short tag on the part of the various \hoot
organizers.

On the client side, our experiments show that a modern computer can
decrypt \msgs significantly faster than Twitter's peak message rate for
the entirety of its traffic. (Even better, the MAC can be validated
before the message ciphertext needs to be decrypted, providing a
shortcut to skip undesired cover traffic.)  Again, computational
overhead isn't going to be the limiting factor.  Instead, the only issue
will be bandwidth.

Group organizers must then take client bandwidth into account when
selecting collisions. An international pop music star might make for an
excellent cover story, but he might simply be too popular for clients,
particularly if the they are using cellular phone networks that haven't
been brought up to the latest multi-megabit speeds.  This is the one
place where the short size of Twitter messages actually works in our
favor. A modest data pipe of 128 kbits/sec can transmit roughly 114
tweets per second. This is well within the regular bounds of most any
short tag selected by the group organizers.

Assuming the entire peak load 7000-or-so tweets per second ends up split
uniformly across short tags with three base-64 digits (i.e., 18 bit
short tags), the average number of tweets per second per short tag is
only 0.02 (i.e., only 1.6 \msgs per minute). That's well within any
realistic bandwidth constraint.

\section{Discussion} \label{sec:discuss}

In this section, we discuss a variety of issues and future extensions of the \hoot design.

\subsection{Deployment}

There are two different paths through which \hoot could be deployed. First, it would be straightforward to implement a proxy service, whether operated by Twitter or by an independent third party, that reads every tweet, encrypts it, and provides a Twitter-like interface to the stream of \msgs. This would incur non-trivial monetary costs for the bandwidth and computation resources, but it is still well within the means of many organizations. A deeper concern is that any country could simply block traffic to or from the \hoot proxy server.

Alternately, the proxy server could read every normal tweet, encrypt it, and post it back to Twitter. This, naturally, would double the number of tweets, which makes it something that Twitter would probably insist on doing themselves rather than accepting from a third-party service. A Twitter-internal implementation might instead, for example, precompute the short tags for every tweet but regenerate \msgs on the fly rather than storing two copies.

Any strategy in which \msgs are injected by a proxy has the property that the \msg need not necessarily specify the original sender's Twitter username.  Breaking the binding between the sender and the public \hoot message would greatly improve sender anonymity, but it would also allow any user who knows the relevant plain tag to impersonate any other member. We could imagine a variety of ways to add some sort of digital signature or hash-chaining layer into the \msg format to differentiate users with ``read-only'' versus ``posting'' privileges. We leave this for future work.

Regardless, a critical question is whether the plain tag for a \msg should ever be sent to Twitter. For users with dedicated clients, the users' plain tags could be stored locally, avoiding any issue where a compromise of Twitter's server becomes a single point of failure for \hoot's confidentiality and integrity. This would clearly be the preferred modality for \hoot usage, but that may not be feasible for a large number of potential \hoot users.

It's important to note that Twitter's web interface offers full SSL encryption. Twitter's dedicated smartphone clients currently use OAuth signatures, protecting message integrity but not privacy. Full encryption would be desirable for smartphone clients to defeat active adversaries who might use deep packet inspection to detect \hoot messages in transit and close the connection.

A few other considerations are worth noting:
\begin{itemize}
\item If Twitter could be convinced to directly support \hoot, then they could certainly stretch the 140-character limit to allow \hoot plaintext messages to be as long as regular Twitter messages.

\item If \hoot users were connected to Twitter's web interface, then it would be essential for \msgs to be decrypted in the browser, via client-side JavaScript, rather than on the server. We want every \msg matching a short tag query to be transmitted over the network to ensure that passive network eavesdroppers, will only see traffic patterns corresponding to those short tags. Sending just the \msgs matching a particular plain tag could leak the group's traffic pattern, even over an SSL-encrypted connection~\cite{liberatore06}.


\item Regardless, an adversary conducting traffic analysis might be able to distinguish a subscription to a \hoot short tag from a subscription to something else, solely based on \msgs' timing and size. This could be addressed by having {\em all} Twitter users receive \hoot cover traffic for a randomly selected \hoot short tag, which their clients would then ignore. It could also be addressed by hacking normal Twitter clients from unsuspecting users to treat any request to follow a hashtag by having them follow the corresponding \hoot short tag, instead.

\item At some point, Twitter may deploy content-aware advertising on its services. Advertising should never be allowed to operate based on a decrypted \hoot, as the advertiser could well be the adversary, and could possibly use analysis of advertisement metrics to violate \hoot users' privacy.

\item Twitter may not have any interest in \hoot or anything like it. See Section~\ref{sec:dht} for a discussion of alternative backend designs.

\end{itemize}

\subsection{Usability}

In essence, \hoot requires its users to manually negotiate a group cryptographic key management system. Even though the use case for \hoot looks and feels much like ``normal'' Twitter hashtags, we are still putting a non-trivial amount of trust into a protocol that requires users to use literal gossip to spread key material. Even highly motivated users could well make mistakes, and any small error would result in the inability to decrypt the desired \msgs. In particular, a significant amount of entropy is required for adequate security, causing \hoot plain tags to push the boundaries on human memorability. In effect, a good \hoot plain tag is comparable to a strong account password. \hoot's use of machine-generated plain tags would have comparable issues with organizations that require users to have strong passwords that are never written down and frequently changed.  And, unlike such organizations which can adopt a variety of ``two factor'' authentication technologies to reduce their reliance on passwords, \hoot fundamentally needs plain tags as strong as strong user passwords.

About the only good news in this process is the growing ubiquity of smartphones, typically having a variety of methods to communicate with other phones nearby. This would allow a set of plain tags to be quickly and painlessly shared via means including two-dimensional barcodes (displayed on one phone's screen, read with another phone's camera), a variety of close-range networking technologies (near-field communication, Bluetooth, ad-hoc WiFi, or infrared file transfer), or even acoustic transfer from one phone's speaker to another phone's microphone. If, in fact, the predominant modality for \hoot plain tag sharing is one of these mechanisms, then plaintag memorizability becomes a non-issue. Plain tags can then be implemented with general-purpose securely-chosen random numbers.

A related usability question is the process for selecting the desired
external hash tag with which to collide a \hoot plain tag. Earlier in
this paper, we cavalierly suggested that famous pop singers helpfully
provide all the cover traffic we might ever want. However, this process
ultimately needs to be handled with care, since recipient deniability
requires the human recipient to have a convincing story under coercive
pressure, and many people may not be able to convincingly demonstrate
their admiration for an overseas pop sensation. A \hoot proxy or other
site could facilitate this by providing information about the popularity
of different short tags, including which ones are trending well in a
given country or region.

\subsection{Adaptability and scalability}

Since \hoot benefits from and encourages collisions in the short tag
space, there will come a point when there are simply too many
collisions, i.e., the ratio of desired messages to cover traffic will
eventually become too small to be practical for the proxy or client
software to decrypt and filter through. The seemingly obvious solution
is to adjust the length of short tags. Adding one character to the tag
length would greatly decrease the chance of random collisions (by
$\frac{1}{|A|}$ for alphabet $A$). However, modifying the tag length
while the system is in use would not be straightforward for
groups. Groups could rely on group leaders to facilitate the process,
e.g. by announcing a new tag to use or giving out multiple plain tags in
advance. Alternatively, the system operators could initially set tags to
be long enough to prevent most random collisions and have groups
emphasize chosen collisions in their plain tag generation.

\subsection{Alternate backends}
\label{sec:dht}

Even though our protocol was designed with Twitter in mind, it is extensible to other systems and platforms. The \hoot protocol describes a secure way to transfer short messages (with low encryption overhead) across virtually any publicly available content distribution network. All that \hoot requires is efficient search primitives for each message's short tag or tags. Everything else is handled by client-side software.

Consequently, if Twitter had no interest in \hoot or concluded that it would not be supportable, then \hoot could just as easily work with a variety of centralized or distributed network services. To pick one possibility, \hoot could use the BitTorrent ``distributed tracker'', or any other large and public distributed hash table (DHT) service, to store messages with the short tags used to name the DHT nodes responsible for their storage. FeedTree~\cite{sandler05feedtree} is one of many systems that have attempted to support micropublishing on a DHT. (Other DHT-based ideas are discussed in Section~\ref{sec:related}.) 

\section{Related Work}
\label{sec:related}

Censorship resistance has been carefully investigated in the context of
publishing documents and
file-sharing~\cite{eternity,freehaven,freenet,gnunet-esed,publius,tangler,dagster}.
Anderson proposed the Eternity Service, which would make documents
available for download and not allow any party to delete any
document~\cite{eternity}. Censorship resistance is provided by
replicating each document over many servers across many legal
jurisdictions. By anonymizing the system's communications, the service
would prevent linking plaintext files to the encrypted versions stored
on the servers by those who did not have the decryption key. Free
Haven~\cite{freehaven}, FreeNet~\cite{freenet}, and
GNUnet~\cite{gnunet-esed} provide similar properties in the context of
peer-to-peer file-sharing, as well as attempting to hide which peers are
hosting which files. Publius~\cite{publius} extends these approaches by
employing Shamir secret sharing~\cite{shamir} to make it harder to
determine what each server is storing. Tangler~\cite{tangler} and
Dagster~\cite{dagster} cryptographically intertwine data from different
documents in such a way that the censor can only force the system to
delete controversial documents by deleting ``legitimate'' documents and
thereby degrading the system as a whole. This provides a
censorship-resistance property similar to one provided by \hoot:
prevention of fine-grained censorship rather than prevention of
heavy-handed censorship.

Censorship resistance has also been studied in the context of
communication more broadly. Some systems aim to evade automated
filters. Feamster et al. propose Infranet, which passes information over
covert channels with the help of participating Web
servers~\cite{infranet}. In another work, Feamster et al. point out that
Infranet and other proxy-based solutions to censorship evasion face the
problem of finding the proxies~\cite{feamster03proxy}. To address this
problem, Feamster et al. require clients to solve cryptographic puzzles
to find a proxy. The Tor anonymity system faces a similar problem. It
has a widely-distributed list of servers~\cite{tor} and thus the censor
could block Tor by blocking access to the servers on the list. Tor uses
{\em bridges}, nodes that allow users to connect to Tor through them, to
evade such blocking~\cite{tor-bridges}. Although the bridges are not
published as widely, they must be disseminated to users and can be
discovered through the same channels by the censors. In particular,
China blocks access to Tor not only through blocking servers but bridges
as well~\cite{tor-china}.

In not relying on proxies, \hoot has the advantage of being easier to
use (since it doesn't require puzzles or CAPTCHAs) and of providing
plausible deniability, whereas the use of proxy-based censorship
resistance tools is likely to be inherently unacceptable to the censor
and could lead to the user being harassed or worse if found out.

\balance


%

\section{Conclusions}

To provide censorship resistance and anonymity for groups wishing to
communicate over a microblogging service like Twitter, we proposed
\hoot. Hoots are private messages that are publicly posted but tagged
with an identifier, allowing interested parties to efficiently find and
decrypt them. By allowing many hash tags to collide with the same
identifier, we protect recipient anonymity and use unrelated traffic as
cover traffic. We found that \hoot can be added to a service like
Twitter with little additional computational resources and reasonable
additional bandwidth costs. We showed that users can have a experience
that's virtually identical to a standard Twitter user, yet with
radically better privacy. We also showed how it would be straightforward
for Twitter to adopt \hoot and deploy it via web interfaces or via
custom clients.

\section{Acknowledgements}

This work was supported in part by the National Science Foundation under
CAREER award number CNS-0954133. Any opinions, findings and conclusions,
or recommendations expressed in this material are those of the author(s)
and do not necessarily reflect those of the National Science Foundation.

\bibliographystyle{abbrv}
\bibliography{hoot}

\balance

\end{document}